\documentclass[prb,twocolumn,showpacs,floatfix,aps]{revtex4}
\usepackage{graphicx}
\newcommand{\etal}{{\it et~al.}~}
\begin{document}
%\preprint{cond-mat/0000000}

\title{Comment on "Experimental determination of superconducting parameters for the
intermetallic perovskite superconductor  MgCNi$_3$" }
\author{Yu.~G.~Naidyuk \footnote{e-mail: naidyuk@ilt.kharkov.ua}}

\affiliation{B.~Verkin Institute for Low Temperature Physics and
Engineering, National Academy  of Sciences of Ukraine, 47 Lenin
Ave., Kharkiv, 61103, Ukraine}

\date{\today}

\begin{abstract}
In a recent paper (Phys.~Rev.~{\bf B 67}, 094502 (2003)) Mao \etal
investigated the bias-dependent conductance of mechanical junctions
between superconducting MgCNi$_3$ and a sharp W tip. They interpreted
their results in terms of 'single-particle tunneling'.
We show it is more likely that current transport through those junctions
is determined by thermal effects due to the huge normal-state resistivity
of MgCNi$_3$.
Therefore no conclusion can be drawn about the possible unconventional
pairing or strong-coupling superconductivity in MgCNi$_3$.
\end{abstract}

\pacs{74.70.Dd, 74.25.Fy, 74.80.Fp, 74.20.Rp}

\maketitle

In a recent paper Mao  \etal  \cite{Mao} reported bulk transport
and specific heat measurements on superconducting MgCNi$_3$.
Additionally, they investigated the conductance of mechanical
junctions between superconducting MgCNi$_3$ and a sharp W tip with
$15\,\mu$m curvature radius. By postulating that tunneling
dominates the conductance they interpreted the observed zero-bias
conductance peak (ZBCP) as caused by Andreev-bound states which
result from a possible unconventional pairing state in MgCNi$_3$.
Besides that, Mao \etal \cite{Mao} attributed the simultaneously
appearing two conductance dips to the characteristic
superconducting energy scale in MgCNi$_3$. On this basis Mao \etal
suggested that 'this result can be taken as further support of
strong-coupling superconductor of MgCNi$_3$'.

We believe that before dealing with more exotic phenomena, like
Andreev-bound states, more trivial effects should be considered to
explain the observed conductance anomalies of those junctions.

\begin{figure}
 \includegraphics[width=8.5cm,angle=0]{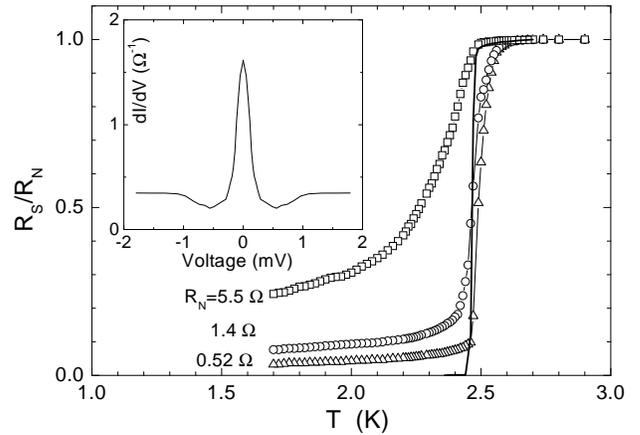}
\caption[]{Temperature dependence of the differential resistance
at zero bias ($R\equiv dV/dI(V=0)$) of three junctions between an
amorphous Zr$_2$Ni ribbon and a Cu tip (open symbols, redrawn from
Ref.\,2). The solid line shows the bulk resistance of the Zr$_2$Ni
ribbon. All curves are normalized to the normal-state resistance
$R_N$, which is also indicated  for each junction. Inset:
Numerically derived from IVC in Ref.~2 differential conductance
$dI/dV(V)$ at 1.7\,K of one of the ZrNi$_2$-Cu contacts.}
\label{zr2ni}
\end{figure}

Firstly, Mao \etal \cite{Mao} assumed that a tunneling barrier
forms at their junctions with rather low resistance $<0.1\Omega$.
This assumption was based only on the fact that the shape of the
superconducting transition in the resistance $R(T)$ of the
junctions deviates from that of the bulk resistivity $\rho (T)$.
According to our experience, such deviations are typical for
mechanical junctions (or point contacts) with highly resistive
metals. The sharp tip damages the sample surface in the contact
area, that means the material there is more degraded than in the
bulk. This can locally change $\rho$ as well as $T_c$.
Fig.\,\ref{zr2ni} shows as example the behavior of $R(T)$ for
several contacts between an amorphous superconducting Zr$_2$Ni
ribbon (its residual normal-state resistivity $\rho_0 \approx
170\,\mu\Omega\,$cm is comparable to that of MgCNi$_3$) and a Cu
tip \cite{Naidyuk1}. While $R(T)$ of low-ohmic contacts has a
rather sharp transition similar to that of $\rho(T)$, the
transition broadens for larger $R_{\rm N}$ The same kind of
broadening was observed for URu$_2$Si$_2$ break junctions
\cite{Naidyuk2} (also a metal with large $\rho$). In the latter
experiments breaking the samples at helium temperatures prevented
the formation of any oxide or other contaminating layer on the
surface which could otherwise produce a tunneling barrier.

Secondly, ZBCPs are characteristic features of junctions formed
with superconductors that have a high normal-state resistivity, as
shown in the inset of Fig.\,\ref{zr2ni} for a ZrNi$_2$ - Cu
contact. As another example, Gloos \etal \cite{Gloos} observed
pronounced zero-bias minima in $dV/dI$ (corresponding to ZBCPs in
$dI/dV$) for contacts between the heavy-fermion superconductor
UBe$_{13}$ and a W tip (UBe$_{13}$ also has a very high
resistivity, comparable to that of MgCNi$_3$). They concluded that
the $dV/dI$ anomalies were due to diffusive and thermal transport
through the junctions, while significant Andreev reflection
currents were missing.

Thirdly, MgCNi$_3$ has a huge residual resistivity $\rho_0 \approx
400\,\mu\Omega \,$cm (see the inset of Fig.\,3 in Ref.\,1) like
that of amorphous metals. With the carrier density $n \approx
10^{28}\,$m$^3$ from Ref.\,6 and the Drude formula $l=\hbar
k_F/(e^2 n \rho)$, where $k_F$ is the Fermi wave number, we
calculate an elastic electron mean free path (mfp) $l_{el} \approx
0.7\,$nm. This is comparable to the lattice constant. The
inelastic mfp $l_{in}\sim \hbar v_F/k_BT\approx 1.5\,\mu$m at 1\,K
according to Ref.\,5. Here $v_F$ is the Fermi velocity. This
results in a very small diffusive inelastic mfp $\Lambda \approx
\sqrt{l_{el}l_{in}} \approx $30\,nm. Applying the Maxwell formula
that describes the spreading resistance of large metallic contacts
\begin{equation}
\label{Rm} R_{\rm N} \approx \rho/2d
\end{equation}
and taking into account that W has a negligibly small resistivity
compared to that in MgCNi$_3$ we estimate a contact radius $r=d/2$
between 12-20$\mu$m for the junctions presented in Ref.\,1. {\it
This fits very well the curvature of the W tip (15\,$\mu$m),
supporting our model of a direct metallic contact!}. Since $d$ is
much larger than the diffusive inelastic electronic mfp $\Lambda$,
these contacts are in the thermal regime\cite{Verkin} in which the
temperature inside the contact rises with applied bias voltage,
and the differential conductance depends only on $\rho(T)$
\cite{Verkin}. In the thermal regime the bias voltage does no
longer determine the excess energy of the electrons. And this
makes it impossible to obtain any spectral information about the
transport processes through the junctions and seriously questions
the conclusions in Ref.\,1 with respect to the characteristic
superconducting energy scale in MgCNi$_3$.

Fourthly, the current density can be quite large for the  contacts
investigated in Ref.\,1. For example at $V = 1\,$mV the  current
density is larger than $j = V/(R_{\rm N}d^2) \simeq
10^7\,$A/m$^2$. The ZBCPs, the abrupt decrease of $dI/dV$ with
increasing bias voltage, are very likely caused by the continuous
growth of the normal phase due to the temperature rise inside the
constriction discussed above as well as due to the increasing
current density. \cite{Iwan} The pulsed-current method used by Mao
\etal \cite{Mao}  to measure the $I-V$ curves with a pulse
duration of $t = 5\cdot 10^{-2}\,$s certainly reduces heating of
the bulk sample itself, but it does not prevent local heating of
the junctions. Junctions with diameters of several $10\,$nm
typically have a thermal relaxation time $\tau \approx
10^{-9}\,$s. \cite{Verkin,Balkashin} Since $\tau \propto d^2$, the
MgCNi$_3$ - W contacts with $d \approx 20-40\,\mu$m, as estimated
above, should have $\tau\sim 10^{-3}\,$s. This is still much
smaller than the pulse duration $t$, meaning that the local
temperature will respond to the applied bias voltage almost
without delay.

In conclusion, to obtain reliable information from point-contact
experiments, the regime of current flow through the constrictions
has to be properly established and/or analyzed. For junctions with
highly resistive metals like MgCNi$_3$ thermal effects have to be
expected. Apparently, they play here the role of preventing us
from energy-resolved spectroscopy.

Discussions with K. Gloos and I. K.Yanson are gratefully
acknowledged.

\end{document}